\documentclass[aip,pop,reprint,superscriptaddress,amsmath,amssymb]{revtex4-2}

\usepackage{physics}
\usepackage{graphicx}
\usepackage{cleveref}
\newcommand{\vk}{{\vb{k}}}
\newcommand{\vp}{{\vb{p}}}
\newcommand{\vq}{{\vb{q}}}

\begin{document}

\title{Three-Wave Forced Hasegawa--Wakatani Model: Reducing Chaos with Zonal Flows}

\author{P. Haerter}
\affiliation{Instituto de Física, Universidade de São Paulo, 05315-970, São Paulo, SP, Brazil.}

\author{R. L. Viana}%
\affiliation{Departamento de Física, Universidade Federal do Paraná, 81531-990, Curitiba,  PR, Brazil.}
\affiliation{Centro Interdisciplinar de Ciência, Tecnologia e Inovação, Núcleo de Modelagem e Computação Científica, Universidade Federal do Paraná, 81531-990, Curitiba,  PR, Brazil}

\author{I. L. Caldas}
\affiliation{Instituto de Física, Universidade de São Paulo, 05315-970, São Paulo, SP, Brazil.}

\begin{abstract}
We investigate the nonlinear dynamics of resistive drift-wave turbulence with a three-wave truncation of the Hasegawa--Wakatani equations, in which the background density gradient either is fixed or evolves self-consistently, fed by an external particle source $P$ (such as neutral beam injection) and eroded by the turbulent particle flux. Two triads that differ only by a small tilt of the long-wavelength mode are compared, in one this mode is zonal and cannot be driven by the gradient, in the other it is a weakly driven drift wave. Period maps in the plane of the adiabaticity parameter $C$ and the drive show that the zonal triad is almost always regular, whereas the non-zonal triad becomes chaotic. The feedback of the flux on the gradient amplifies this contrast: the fraction of aperiodic states grows considerably in the non-zonal triad and remains negligible in the zonal one. Along a line of constant $C$ at low adiabaticity, the non-zonal triad with a fixed gradient enters chaos through torus breakdown with intermittency and leaves it through an interior crisis and a period-doubling cascade. With the self-consistent gradient, a Neimark--Sacker bifurcation leads to a phase-locked torus that breaks down abruptly, with hysteresis, and chaos then persists up to the largest drive, with the gradient driven intermittently above the threshold of the long-wavelength mode. None of these routes to chaos exists when that mode is zonal.
\end{abstract}

\maketitle

\section{Introduction}

One of the key problems in toroidal magnetic confinement systems is the underlying mechanism behind the emergence and suppression of anomalous transport\cite{hortonDriftWavesTransport1999,balescuAspectsAnomalousTransport2005,hortonTurbulentTransportMagnetized2017}. By understanding how turbulence, or its components, arises in magnetically confined plasmas, one can use methods to control it and obtain information about how the transport will behave\cite{balescuAspectsAnomalousTransport2005}. Such turbulent processes often display a broad fluctuation spectrum, with maxima at the longest measured scales (small wave vectors) and low frequencies\cite{hortonDriftWavesTransport1999}. Because wave-like instabilities occur only over a finite range of the spectrum, nonlinear mechanisms such as three- or four-wave interactions are essential for explaining the broadband fluctuations observed in laboratory plasmas\cite{terryStochasticityRandomPhase1982,batistaNonlinearThreemodeInteraction2006,kimMeasurementsNonlinearEnergy1997}.

Among the main waves that affect such systems are the drift waves, which arise from the coupling between oscillations of the electrostatic potential and the density in the presence of a background density gradient, and which play an important role in transport physics\cite{hortonDriftWavesTransport1999,hortonNonlinearDriftWaves1990}. In high-temperature plasmas, sufficiently steep density gradients at the plasma edge can give rise to fully developed drift-wave turbulence, which is considered a likely candidate for explaining the observed anomalous transport rates\cite{balescuAspectsAnomalousTransport2005,diamondZonalFlowsPlasma2005}.

Many models have been proposed to describe the emergence and evolution of this type of turbulence in magnetically confined plasmas. One of the first and most influential is the Hasegawa--Mima (HM) model\cite{hasegawaPseudothreedimensionalTurbulenceMagnetized1978,hasegawaStationarySpectrumStrong1977}, which captures the nonlinear dynamics of the electrostatic potential in a magnetized plasma. It rests on simple assumptions, quasineutrality, a uniform and constant magnetic field, cold ions responding through the polarization drift, and Boltzmann (adiabatic) electrons. Even so, it is a robust model, widely used to describe drift-wave propagation and the formation of zonal flows\cite{chandreHamiltonianFormulationModified2014,satoGeneralizedHasegawaMimaEquation2022,satoHamiltonianStructureNonlinear2024}.

The main limitation of the HM model is its absence of linear instability, where the density and potential fluctuations are exactly in phase, so there is no cross-field particle flux, and the system needs external parameters and forcing to display behaviors that are normally observed in confined plasmas\cite{terryStochasticityRandomPhase1982,hortonQuasitwodimensionalDynamicsPlasmas1994}. To overcome this, Hasegawa and Wakatani proposed a more complete model\cite{hasegawaPlasmaEdgeTurbulence1983} (HW), in which the plasma density is incorporated into the dynamics and the potential and density are coupled through an adiabaticity parameter, which measures the strength of the parallel electron response, making each respond to changes in the other\cite{wakataniCollisionalDriftWave1984,hasegawaPlasmaEdgeTurbulence1983}. This makes the HW model a minimal model of drift-wave turbulence, owing to the presence of linear instability, nonlinear resonant interactions, and zonal flows.

Zonal flows are axisymmetric sheared $\mathbf{E}\times\mathbf{B}$ flows, generated by the turbulence itself through nonlinear interactions, that regulate transport by shearing the turbulent eddies\cite{diamondZonalFlowsPlasma2005,kimZonalFlowsTransient2003,scottEnergeticsInteractionElectromagnetic2005,schmitzRoleZonalFlow2012,xuBlobHoleFormation2009,zhangInfluenceZonalFlow2020}. In the HW system they have a distinctive property, since they have no variation along the drift direction ($k_y=0$), the density gradient cannot drive them, and they can only exchange energy with the drift waves through nonlinear coupling. They therefore act as a pure regulator of the turbulence, rather than as one more unstable wave\cite{gurcanPhaseAmplitudeEvolution2022,guillonPhaseTransitionTurbulence2025}.

In the HW model the density is a dynamical field, but it evolves on top of a background density gradient that supplies the free energy for the instability. This gradient normally corresponds to a stationary profile that is kept constant over the plasma by external means. This is a strong simplification, since in a real plasma the profile is set by the balance between the particle sources and the turbulent flux, and the turbulence, in turn, relaxes the very gradient that drives it\cite{numataBifurcationElectrostaticResistive2007,konigesStructureFormationTransport1992}. This feedback is the basis of predator--prey descriptions of drift-wave turbulence, in which the gradient plays the role of the prey, fed by an external source, and the drift waves play the role of the predators, growing by consuming it, so that the two influence each other and give rise to regular or chaotic oscillations\cite{diamondSelfRegulatingShearFlow1994,reiserPlasmaSourceDriven2014,schmitzRoleZonalFlow2012}.

A natural way to study these ideas is through low-dimensional truncations. Keeping only a few Fourier modes that satisfy the resonance condition $\mathbf{k}_1+\mathbf{k}_2+\mathbf{k}_3=0$ yields a triad, the elementary unit of nonlinear energy transfer, whose dynamics is described by a small set of ordinary differential equations\cite{batistaNonlinearThreemodeInteraction2006,terryStochasticityRandomPhase1982,gurcanPhaseAmplitudeEvolution2022}. Despite its simplicity, such a system allows tools that are not suitable for full turbulence simulations, such as bifurcation analysis, Lyapunov exponents, and period maps. In the adiabatic limit the HW triad reduces to the three-wave truncation of the HM equation, whose dynamics is well characterized. Keeping the first correction beyond this limit leads to a triad with growth and damping rates, which has been studied with these rates introduced phenomenologically. In the HW system, these rates follow directly from the resistive coupling and the density gradient, and the effect of letting the gradient evolve self-consistently, together with the zonal character of the long-wavelength mode, has not been addressed.

In this work we couple the HW triad to a time-dependent density gradient, which evolves self-consistently, fed by a particle source and eroded by the turbulent particle flux. To isolate the role of the zonal flow, we compare two triads that differ only by a small tilt of the long-wavelength mode. In one triad, it is purely zonal and cannot be driven by the density gradient, while in the other it is slightly tilted off the zonal axis and behaves as a weakly driven drift wave. Each triad is studied with both a fixed and a self-consistent gradient. Our main finding is that the zonal triad is almost always regular, whereas the non-zonal triad becomes chaotic over a sizeable region of the parameter space, and that the feedback of the flux on the gradient enlarges this region. We also identify the routes to chaos of the non-zonal triad, none of which exists when the long-wavelength mode is zonal, so the zonal flow acts as a mechanism that reduces chaos.

In Section~\ref{sec:Model} the Hasegawa--Wakatani model is presented and its wave expansion is obtained, together with the inclusion of the non-constant density background, and the linear stability of the modes is explored. Then, in Section~\ref{sec:parameters}, the constants and parameters used are presented, together with the zonal and non-zonal triads and the numerical methods. Later, in Section~\ref{sec:fixed}, a reference case is defined and explored, with the background density gradient kept constant, comparing the two triads. In Section~\ref{sec:dynamic} the background density gradient is made non-stationary, which allows oscillations and chaotic and periodic regions intermixed in parameter space, and the comparison between the triads is repeated. Section~\ref{sec:routes} then explores the routes to chaos of the non-zonal triad, which are absent when the long-wavelength mode is zonal. Finally, in Section~\ref{sec:conclusions}, the conclusions of the work are presented.

\section{Model}
\label{sec:Model}
\subsection{Hasegawa--Wakatani equations}

The Hasegawa--Wakatani (HW) model\cite{hasegawaPlasmaEdgeTurbulence1983,wakataniCollisionalDriftWave1984,hasegawaSelforganizationElectrostaticTurbulence1987} describes the edge turbulence of a plasma, where a steady background density in the radial direction is normally assumed, and the fluctuations lie in the plane $x$--$y$ perpendicular to a uniform and constant magnetic field $\vb{B}=B\hat{z}$. The spatial variable $x$ stands for the radial direction and $y$ for the poloidal direction. Lengths are normalized to the ion sound gyroradius $\rho_s=c_s/\omega_{ci}$, where $c_s=\sqrt{T_e/m_i}$ is the ion sound speed, $T_e$ the electron temperature (in energy units) and $m_i$ the ion mass, and time $t$ is normalized to the inverse ion cyclotron frequency $\omega_{ci}^{-1}$, with $\omega_{ci}=eB/m_i$. It is an improvement on the Hasegawa--Mima equation: besides the fluctuation of the electrostatic potential $\phi$, it also includes the variation of the plasma density $n$, which allows a better development of the turbulence and a more precise description of the interaction between the potential and the ions.  The electrostatic potential $\phi$ and the density fluctuation $n$ obey

\begin{align}
    &\pdv{t}\nabla^2\phi + (\hat{z}\cross\nabla\phi)\cdot\nabla\nabla^2\phi = C(\phi-n)+\nu\nabla^4\phi\\
    &\pdv{t} n + (\hat{z}\cross\nabla\phi)\cdot\nabla n + \kappa\pdv{\phi}{y} = C(\phi-n)+D\nabla^2n
\end{align}

 Here $C$ is the adiabaticity parameter, defined below, and $\nu$ and $D$ are the viscosity and the particle diffusion coefficients, which damp the vorticity and the density fluctuations, respectively. These equations display a similar structure to the HM equation, with $\nabla^2\phi$ being the vorticity of the plasma, and the nonlinear interactions are given by $ (\hat{z}\cross\nabla\phi)\cdot\nabla$ operator, describing the advection of the multiplied quantity by the $E\cp B$ drift velocity, with $\hat{z}$ being the unit vector parallel to the magnetic field. The background density gradient $\kappa\equiv-\dv*{\ln n_0}{x}$ drives the instability through the term $\kappa\,\partial_y\phi$.

The coupling between the potential and the density are given by the adiabaticity parameter $C$, that can be written as 

\begin{equation}
    C=\frac{T_e\,k_\parallel^2}{e^2 n_0\,\eta_\parallel\,\omega_{ci}}
\end{equation}
measures the parallel electron response, where $k_\parallel$ is the parallel wave number and $\eta_\parallel$ the parallel resistivity. When $C\rightarrow0$, the potential and the density decouple and the Navier--Stokes equation is recovered, being called the hydrodynamic regime. On the other hand, in the adiabatic limit, when $C\rightarrow\infty$, the coupling term dominates and $\tilde{\phi}\simeq\tilde{n}$ at the leading order, then recovering the Hasegawa-Mima equation. This case is normally associated with the electron parallel resistivity going to zero, causing the perturbation of the electrostatic potential to be in the phase with the fluctuation density. There, the system only presents drift waves propagating in the $y$ direction, and there is no linear instability to drive turbulence, since density and potential are in phase and the cross-field particle flux vanishes. 

To achieve a more realistic construction of the model, we also include some small scale dissipation terms on the fluctuations through the viscosity coefficients $\nu$ and $D$, acting, respectively, on vorticity $\nabla ^2 \phi$ and the density. These diffusion terms are mainly used for numerical purpose, to balance the exponential energy injection by the linear instability, and avoid the accumulation of energy which reaches the smallest scales by the direct enstrophy cascade. 

\subsection{Non constant density Gradient}

The gradient $\kappa$ is usually taken as a constant, which corresponds to a background profile maintained by external means. But to achieve a more realistic turbulence response, we can consider that the background density gradient is a function of time $S(t)$, that is connected to already known predator--prey models in plasma turbulence \cite{schmitzRoleZonalFlow2012,kimZonalFlowsTransient2003}, considering it at the form 

\begin{equation}
    \dv{S}{t} = P - \lambda S(t) - \Gamma_k
    \label{eq:S_t}
\end{equation}

where there is a pumping intensity $P$ that can be related to an external constant forcing, such as constant beam injection, or a constant depletion of ions into the wall. The rate $\lambda$ describes the relaxation of the gradient by transport channels other than the turbulent flux, such as collisional diffusion. The last term is the turbulent radial particle flux  averaged over the entire $2D$ space,  

\begin{equation}
    \Gamma_k \equiv \langle n v_x \rangle_{x,y}
    \label{eq:Gamma_k}
\end{equation}
where $v_x=-\partial_y\phi$ is the radial $\vb{E}\cross\vb{B}$ velocity. Without fluctuations, $S$ relaxes to $S_0=P/\lambda$, the gradient set by the source alone and $\langle \cdot\rangle_{x,y}$ denotes averaging over both $x$ and $y$ directions. Using the Fourier decomposition of both fields, and summing over one half of the wave-vector plane, the modes $\vk$ and $-\vk$ contribute equally, we can re-write this as 

\begin{equation}
    \Gamma_k = -2\sum_{k_x,k_y}k_y\Im{n_k\phi^*_k}
    \label{eq:Final_gamma}
\end{equation}

Then, with the new background density, the set of equations for the Hasegawa--Wakatani model becomes

\begin{align}
    &\pdv{t}\nabla^2\phi + (\hat{z}\cross\nabla\phi)\cdot\nabla\nabla^2\phi = C(\phi-n)+\nu\nabla^4\phi\\
    &\pdv{t} n + (\hat{z}\cross\nabla\phi)\cdot\nabla n + S(t)\pdv{\phi}{y} = C(\phi-n)+D\nabla^2n\\
    &\dv{S}{t} = P - \lambda S(t) + 2\sum_k k_y\Im{\phi_k^*n_k}
\end{align}

\subsection{Wave Modes}

To study the nonlinear interactions and energy transfer within the system, we can expand both $\phi$ and $n$ in Fourier wave modes $\vk$, thus allowing us to simplify the nonlinear operator $ (\hat{z}\cross\nabla\phi)\cdot\nabla$ to some constant multiplication in the wave space. 

\begin{equation}
     \phi = \sum_\vk \phi_\vk(t)e^{i\vk\cdot\vb{r}}\qquad      n = \sum_\vk n_\vk(t)e^{i\vk\cdot\vb{r}} 
\end{equation}
with $\phi_{-\vk}=\phi_\vk^*$ and $n_{-\vk}=n_\vk^*$, the HW equations become
\begin{align}
     \dv{\phi_\vk}{t} +(A_k-B_k)\phi_\vk      &= \frac{C}{k^2} n_\vk(t)+\alpha_\vk^{\vp,\vq}\\
    \dv{n_\vk}{t} +(A_k+B_k)n_\vk      &= (C-iS(t)k_y)\phi_k +\gamma_\vk^{\vp,\vq}
\end{align}
with the nonlinear terms
\begin{align}
    \alpha_\vk^{\vp,\vq} &= \frac{1}{2}\sum_{\vp+\vq=\vk}\frac{(q^2-p^2)}{k^2}\qty(\vp\cross \vq)\cdot   \hat{z} \phi_\vq\phi_\vp\\
    \gamma_\vk^{\vp,\vq} &=  \sum_{\vp+\vq=\vk}\qty(\vp\cross\vq )\cdot\hat{z}  \phi_\vp n_\vq
\end{align}
and the linear coefficients
\begin{align}
    A_k &= \frac{1}{2}\qty[(Dk^2+C)+\qty(\frac{C}{k^2}+\nu k^2)]\\
    B_k &= \frac{1}{2}\qty[(Dk^2+C)-\qty(\frac{C}{k^2}+\nu k^2)] 
\end{align}

To analyze the fundamental resonant interactions driving the turbulence, we truncate this infinite-dimensional system to consider only three coupled interacting waves that satisfy the triad resonance condition $\vk_1 + \vk_2 + \vk_3=0$\cite{batistaNonlinearThreemodeInteraction2006}. Defining the geometric interaction coefficient $V$ as

\begin{equation}
    V = (\vb{k}_1\cross\vb{k}_2)\cdot \hat{z}= (\vb{k}_2\cross\vb{k}_3)\cdot \hat{z}= (\vb{k}_3\cross\vb{k}_1)\cdot \hat{z}
\end{equation}

The truncated system evaluates to the following set of ordinary differential equations 

\begin{align}
    &\dv{\phi_1}{t} + (A_1-B_1)\phi_1 = \frac{C}{k_1^2}n_{1} + \frac{V}{k_1^2}(k_3^2-k_2^2)\phi_2^*\phi_3^*,\label{eq:triad_first}\\
    &\dv{\phi_2}{t} + (A_2-B_2)\phi_2 = \frac{C}{k_2^2}n_{2} + \frac{V}{k_2^2}(k_1^2-k_3^2)\phi_3^*\phi_1^*,\\
    &\dv{\phi_3}{t} + (A_3-B_3)\phi_3 = \frac{C}{k_3^2}n_{3} + \frac{V}{k_3^2}(k_2^2-k_1^2)\phi_1^*\phi_2^*,\\
    &\dv{n_1}{t} + (A_1+B_1)n_1 = (C-iSk_{1y})\phi_1 +  V(\phi_2^*n_3^*-\phi_3^*n_2^*),\\
    &\dv{n_2}{t} + (A_2+B_2)n_2 = (C-iSk_{2y})\phi_2 +  V(\phi_3^*n_1^*-\phi_1^*n_3^*),\\
    &\dv{n_3}{t} + (A_3+B_3)n_3 = (C-iSk_{3y})\phi_3 +  V(\phi_1^*n_2^*-\phi_2^*n_1^*),\\
    &\dv{S}{t} = P - \lambda S + 2\sum_{j=1}^{3}k_{jy}\Im\{\phi_j^*n_j\}.\label{eq:triad_last}
\end{align}

Within this three-wave approximation, the Fourier expansion of the electrostatic potential is given by

\begin{align}
    \phi(\vb{x},t) = \sum_{j=1}^{3}\qty[\phi_j(t)e^{i\vb{k}_j\cdot\vb{x}}+\mathrm{c.c.}]\\
    n(\vb{x},t) = \sum_{j=1}^{3}\qty[n_j(t)e^{i\vb{k}_j\cdot\vb{x}}+\mathrm{c.c.}]
\end{align}

Equations~\eqref{eq:triad_first}--\eqref{eq:triad_last} define a 13-dimensional real system: three complex amplitudes $\phi_j$, three complex amplitudes $n_j$, and the real gradient $S$. The dissipation removes any conserved quantity. The system is, however, invariant under the phase rotations
\begin{equation}
    \phi_j\to\phi_j e^{i\theta_j},\qquad n_j\to n_j e^{i\theta_j},\qquad \theta_1+\theta_2+\theta_3=0,
    \label{eq:symmetry}
\end{equation}
which leave the quadratic couplings and the flux unchanged. This two-parameter symmetry reduces the dynamics to eleven phase-invariant variables: the six moduli $|\phi_j|$ and $|n_j|$, the three phase differences $\arg n_j-\arg\phi_j$, the total phase $\Theta=\arg\phi_1+\arg\phi_2+\arg\phi_3$, and $S$.

This truncated three-wave system forms the theoretical basis for our dynamical analysis. By systematically varying the parameter space for the coupling strength $C$ and the energy pump drive $P$, we can simulate how the NBI (Neutral Beam Injection) energy pump impacts mode stability and induces structural bifurcations within the plasma edge.

\subsection{Linear Stability Analysis}
\label{sec:linear}

The truncated system admits the trivial equilibrium
\begin{equation}
    \phi_j = n_j = 0, \qquad S = S_0 \equiv \frac{P}{\lambda}, \qquad j=1,2,3,
    \label{eq:trivial_eq}
\end{equation}
in which the gradient is set only by the balance between the source $P$ and the relaxation $\lambda S$. To study its stability we add small perturbations $\phi_j$, $n_j$ and $S=S_0+\delta S$, and keep only the terms that are linear in them. The triad couplings and the particle flux are products of two fluctuations, so they are of second order and drop out. The linearized system is
\begin{align}
    &\dv{\delta S}{t} = -\lambda\,\delta S, \label{eq:lin_S}\\
    &\dv{\phi_j}{t} = -(A_j-B_j)\,\phi_j + \frac{C}{k_j^2}\,n_j, \label{eq:lin_phi}\\
    &\dv{n_j}{t} = \qty(C - iS_0k_{jy})\,\phi_j - (A_j+B_j)\,n_j. \label{eq:lin_n}
\end{align}
The perturbation of the gradient always decays at rate $\lambda$, and the three modes evolve independently of each other. At linear order they only share the value of $S_0$, and the coupling between modes is purely nonlinear. The stability of the fixed point is therefore decided mode by mode, by the pair $(\phi_j,n_j)$ and the diagonal coefficients are given by
\begin{equation}
    A_j - B_j = \frac{C}{k_j^2} + \nu k_j^2, \qquad A_j + B_j = C + D k_j^2,
\end{equation}
are the damping rates of $\phi_j$ and $n_j$ when the two fields are not coupled, so $A_j$ is the mean of the two damping rates and $B_j$ is half their difference. In the rest of this section we drop the index $j$.

Equations~\eqref{eq:lin_phi} and \eqref{eq:lin_n} form a linear system with constant coefficients,
\begin{equation}
    \dv{t}\begin{pmatrix}\phi\\ n\end{pmatrix}
    = \mathsf{M}\begin{pmatrix}\phi\\ n\end{pmatrix},
    \qquad
    \mathsf{M} = \begin{pmatrix} -(A-B) & C/k^2\\ C - iS_0k_y & -(A+B)\end{pmatrix}.
    \label{eq:lin_modes}
\end{equation}
Substituting $(\phi,n)=(\hat\phi,\hat n)\,e^{\sigma t}$, with $\hat\phi$ and $\hat n$ constant, gives $\mathsf{M}(\hat\phi,\hat n)^T=\sigma(\hat\phi,\hat n)^T$, so $\sigma$ is an eigenvalue of $\mathsf{M}$. A nonzero solution exists only if $\det(\mathsf{M}-\sigma\mathsf{I})=0$, that is,
\begin{equation}
    (\sigma + A - B)(\sigma + A + B) - \frac{C}{k^2}\qty(C - iS_0k_y) = 0.
    \label{eq:charpoly}
\end{equation}
The first product is $(\sigma+A)^2-B^2$, so this quadratic equation has the two roots
\begin{equation}
    \sigma = -A \pm \sqrt{\Delta}, \qquad
    \Delta = B^2 + \frac{C^2}{k^2} - i\,\frac{C S_0 k_y}{k^2}.
    \label{eq:sigma}
\end{equation}
Because of the drive term, $\mathsf{M}$ is complex and so is $\sigma$. We write $\sigma=\gamma-i\omega$, so that the perturbation behaves as $e^{\gamma t}e^{-i\omega t}$. Together with the spatial factor $e^{i\vk\cdot\vb{x}}$, each mode is a wave $e^{\gamma t}e^{i(\vk\cdot\vb{x}-\omega t)}$ with frequency $\omega$ and growth rate $\gamma$, and it is unstable when $\gamma>0$.

To separate $\gamma$ and $\omega$ we split $\Delta$ into its real and imaginary parts. From the definitions of $A$ and $B$,
\begin{align}
    A^2 - B^2 &= (A-B)(A+B) = \frac{C^2}{k^2} + G,\\
    G &\equiv CD + \nu C k^2 + \nu D k^4 \geq 0,
    \label{eq:Gk}
\end{align}
where $G$ collects all the dissipative terms. Then $\Delta = X - iY$, where
\begin{equation}
    X = A^2 - G, \qquad Y = \frac{C S_0 k_y}{k^2}
\end{equation}
are real. $X$ does not depend on the gradient, while $Y$ is proportional to it.

The growth rate and the frequency are the real and imaginary parts of $\sigma=-A\pm\sqrt{\Delta}$, so we need the square root of the complex number $\Delta$. We write it as $\sqrt{\Delta}=a-ib$, with $a$ and $b$ real and $a\geq0$. Squaring and comparing with $\Delta=X-iY$ gives
\begin{equation}
    a^2-b^2 = X, \qquad 2ab = Y.
\end{equation}
Substituting $b=Y/2a$ in the first equation gives a quadratic equation for $a^2$, $a^4 - Xa^2 - Y^2/4=0$, whose positive solution is
\begin{equation}
    a = \sqrt{\frac{X+\sqrt{X^2+Y^2}}{2}}, \qquad b=\frac{Y}{2a}.
\end{equation}
The two roots are therefore $\sigma=-A+a-ib$ and $\sigma=-A-a+ib$. The second one has $\gamma=-A-a<0$ and is always damped. The first one is the drift wave, with growth rate and frequency
\begin{equation}
    \gamma = -A + a, \qquad \omega = b = \frac{Y}{2a}.
    \label{eq:gamma}
\end{equation}
Since $a>0$, the frequency has the sign of $Y$, that is, of $S_0k_y$.

The mode grows when $a>A$. Without a gradient, $S_0=0$, we have $Y=0$ and $a=\sqrt{X}$. Since $X=A^2-G$ and $G\geq0$, this gives $a\leq A$, so every mode is damped, as expected in the absence of an energy source. As the gradient increases, $|Y|$ increases, and so does $a$, since $Y$ enters only through $\sqrt{X^2+Y^2}$. The growth rate therefore increases with the drive $|S_0k_y|$, and the mode becomes unstable when $a$ reaches $A$. Setting $a=A$ in the expression for $a$ gives
\begin{equation}
    X + \sqrt{X^2+Y^2} = 2A^2 .
\end{equation}
Isolating the square root and squaring, the terms $X^2$ cancel, and
\begin{equation}
    Y^2 = 4A^2\,(A^2-X) = 4A^2 G .
\end{equation}
With $Y=CS_0k_y/k^2$, mode $\vk$ is unstable when $|S_0|$ exceeds the threshold
\begin{equation}
    S_c(\vk) = \frac{2A k^2}{C\,|k_y|}\sqrt{G}, \qquad G = CD + \nu Ck^2 + \nu D k^4 .
    \label{eq:Sc}
\end{equation}

The threshold is the ratio between the dissipation, contained in $G$, and the efficiency $C|k_y|/k^2$ with which the gradient drives the mode. In the ideal limit $D=\nu=0$ we have $G=0$ and $S_c=0$, which recovers the classical result that any gradient destabilizes the drift wave. The factor $C$ in the denominator shows that the drive requires the coupling between density and potential, and for $C=0$ no mode can be destabilized. Modes with $k_y=0$ have $S_c\to\infty$ and are never driven, since the drive term $S\,\partial\phi/\partial y$ vanishes. This also holds in the modified model used for the zonal mode (Sec.~\ref{sec:parameters}), where the coupling $C$ is removed for $k_y=0$. In that case $\mathsf{M}$ is diagonal, with entries $-\nu k^2$ and $-Dk^2$, and the mode is simply damped.

At the threshold, $a=A$ and $S_0=S_c$, and Eq.~\eqref{eq:gamma} gives the frequency of the marginal wave,
\begin{equation}
    \omega_c = \frac{C\,S_c\,k_y}{2A k^2}.
    \label{eq:omega_c}
\end{equation}

In the triad, each mode has its own threshold, $S_c(\vk_1)$, $S_c(\vk_2)$ and $S_c(\vk_3)$, and the fixed point of Eq.~\eqref{eq:trivial_eq} becomes unstable as soon as one of them is exceeded. We call the lowest of the three thresholds $S_\mathrm{th}$, the threshold of the triad. Which mode sets it depends on the triad geometry and on $C$ (Sec.~\ref{sec:parameters}). Since $S_0=P/\lambda$, the fixed point loses stability when the source exceeds
\begin{equation}
    P_c = \lambda\, S_\mathrm{th}.
    \label{eq:Pc}
\end{equation}
Both $S_\mathrm{th}$ and $P_c$ depend on $C$ through $A$ and $G$, and the curve $P_c(C)$ is the linear stability boundary in the $(C,P)$ plane.

\begin{figure*}
    \centering
    \includegraphics[width=\textwidth]{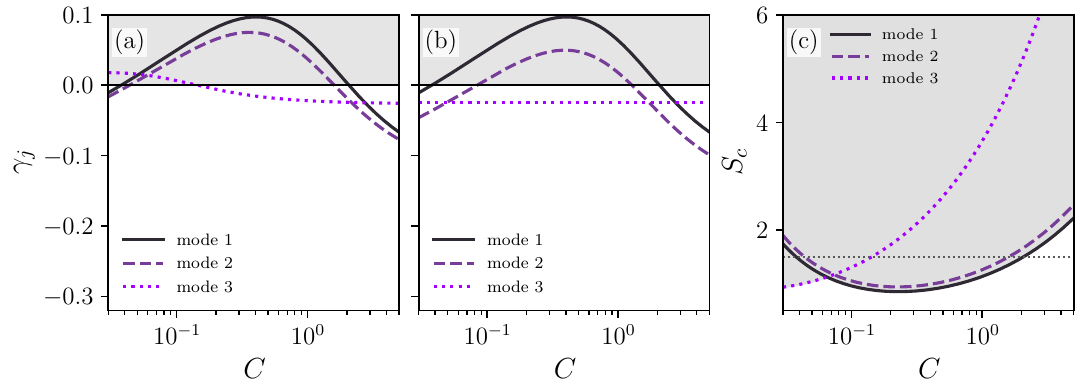}
    \caption{Linear stability of the trivial equilibrium for $D=\nu=0.1$. Growth rate $\gamma_j=\Re\,\sigma_j$ of the drift-wave root of Eq.~\eqref{eq:sigma} for each mode $j$ as a function of $C$, at fixed $S_0=1.5$, for (a) the non-zonal triad and (b) the zonal triad. The shaded band marks $\gamma>0$ (instability). In the zonal triad, mode 3 has $k_y=0$ and the adiabatic coupling is removed for it, so it is damped at the rate $\nu k_3^2$ for every $C$. (c) Marginal stability curves $S_c(\vk_j)$, Eq.~\eqref{eq:Sc}, for the modes of the non-zonal triad. A mode is unstable above its curve, and the shaded region is unstable for the triad. The dotted horizontal line marks $S_0=1.5$, the value used in (a) and (b). The threshold of the zonal triad is that of mode 1, and it coincides with the threshold of the non-zonal triad for $C\gtrsim0.065$.}
    \label{fig:linear}
\end{figure*}

Figure~\ref{fig:linear} shows these results for the parameters of Sec.~\ref{sec:parameters}. Panels (a) and (b) give the growth rate $\gamma$ of each mode as a function of $C$ at a fixed gradient, $S_0=1.5$. A mode is unstable where its curve lies above zero and stable where it lies below. Each drift wave is unstable only within a window of $C$. For small $C$ the drive, which is proportional to $C$, vanishes and the dissipation dominates. For large $C$ the fields approach the adiabatic response $n\approx\phi$, the flux that sustains the instability vanishes, and the mode is again damped. In between, $\gamma$ has a maximum, near $C\approx0.4$ for modes 1 and 2. In the non-zonal triad, mode 1 is unstable for $0.038<C<2.08$ and mode 2 for $0.044<C<1.60$. In the zonal triad mode 1 is the same, while mode 2 is unstable for $0.083<C<1.28$, because $k_{2y}$ differs. Mode 3 behaves differently in the two triads. In the zonal triad it is damped for every $C$, at the rate $\nu k_3^2=0.025$. In the non-zonal triad it is weakly driven and is unstable only for $C<0.144$ at this $S_0$.

Panel (c) shows the same information in the $(C,S_0)$ plane, through the marginal curves $S_c(\vk_j)$ of Eq.~\eqref{eq:Sc}. A mode is unstable for $S_0$ above its curve, so the unstable region of the triad lies above the lowest curve, $S_\mathrm{th}(C)$. The horizontal line $S_0=1.5$ crosses the curves at the edges of the windows in panels (a) and (b). The threshold has a minimum, $S_c\approx0.85$ at $C\approx0.23$, set by mode 1, so for $S_0$ below this value the fixed point is stable for every $C$. The thresholds of modes 1 and 3 cross at $C\approx0.065$. For smaller $C$ the long-wavelength mode of the non-zonal triad is the first to become unstable, and its threshold falls below that of mode 1, which sets the threshold of the zonal triad.

We now compute the particle flux carried by a linear mode, which enters the equation for $S$. In a linear mode, $n$ and $\phi$ are not independent. Setting $\dv*{\phi}{t}=\sigma\phi$ in Eq.~\eqref{eq:lin_phi} gives
\begin{equation}
    n = \frac{k^2}{C}\,(\sigma + A - B)\,\phi ,
\end{equation}
so the density is the potential multiplied by a complex factor, which sets the phase shift between the two fields. Multiplying by $\phi^*$ and taking the imaginary part, only $\Im\sigma=-\omega$ survives, since $A-B$ is real, and
\begin{equation}
    \Im\{\phi^* n\} = -\frac{k^2}{C}\,\omega\,|\phi|^2 .
\end{equation}
The contribution of mode $j$ to the flux in Eq.~\eqref{eq:S_t} is then
\begin{equation}
    \Gamma_j = -2k_{jy}\Im\{\phi_j^* n_j\} = \frac{2k_{jy}k_j^2}{C}\,\omega_j\,|\phi_j|^2 .
    \label{eq:Gamma_j}
\end{equation}
Since $\omega_j$ has the sign of $S_0k_{jy}$, the product $k_{jy}\omega_j$ is positive for $S_0>0$, and so is $\Gamma_j$. The flux therefore always reduces $S$. The unstable wave carries particles down the gradient and weakens the gradient that drives it, which is the predator--prey mechanism of the model. This flux exists only because $n$ and $\phi$ are out of phase. For the adiabatic response $n=\phi$, $\Im\{\phi^*n\}=0$ and there is no transport.

This feedback also determines what happens just above the threshold. Consider a state in which only one mode $j$ is excited. All triad couplings vanish, since each of them is a product of two different modes, so the mode remains alone for all times. It obeys Eqs.~\eqref{eq:lin_phi} and \eqref{eq:lin_n} with $S(t)$ in place of $S_0$, while $S$ obeys Eq.~\eqref{eq:S_t} with the flux $\Gamma_j$. The gradient is then pushed towards the threshold of the mode from both sides. If $S>S_c(\vk_j)$, the mode grows, the flux increases and $S$ decreases. If $S<S_c(\vk_j)$, the mode decays, the flux decreases and the source increases $S$. The system reaches a steady state with
\begin{equation}
    S = S_c(\vk_j),
\end{equation}
in which the mode neither grows nor decays, and rotates at the frequency $\omega_c$ with constant amplitude. The amplitude follows from $\dv*{S}{t}=0$ in Eq.~\eqref{eq:S_t}, that is, $P-\lambda S_c=\Gamma_j$. With $\omega_j=\omega_c$, Eqs.~\eqref{eq:omega_c} and \eqref{eq:Gamma_j} give $\Gamma_j=k_{jy}^2S_c|\phi_j|^2/A_j$, and therefore
\begin{equation}
    |\phi_j|^2 = \frac{A_j\,\qty[P - \lambda S_c(\vk_j)]}{k_{jy}^2\, S_c(\vk_j)}.
    \label{eq:single_mode}
\end{equation}

This result has three consequences. First, for the mode with the lowest threshold, $\lambda S_c(\vk_j)=P_c$, the amplitude is zero at $P=P_c$ and grows as $\sqrt{P-P_c}$. Second, in this single-mode state only the phases rotate, as allowed by the phase symmetry of Eq.~\eqref{eq:symmetry}, while all amplitudes and $S$ are constant. In the period maps, which are based on amplitudes, this orbit appears as an equilibrium. Third, above the threshold the gradient no longer increases with $P$. It remains clamped at $S_c$, and the additional source is removed by the flux, through the growth of $|\phi_j|$. This state exists only with a self-consistent gradient. For a fixed $S$, nothing limits the growth of the unstable mode except the triad coupling. The stability of the single-mode state, and the dynamics that appears when it is lost, are discussed in Sec.~\ref{sec:dynamic}.

\section{Parameters and numerical methods}
\label{sec:parameters}
In order to explore the effects of the change from a constant to a time dependent background density gradient, we need to set up some numerical parameters. 

All quantities are in the standard Hasegawa--Wakatani normalization, with lengths in units of the ion sound gyroradius $\rho_s$ and time in units of the inverse ion cyclotron frequency $\omega_{ci}^{-1}$. In these units, $k\sim1$ corresponds to the drift-wave scale $k\rho_s\sim1$, where the linear drive is strongest.

The dissipation coefficients are fixed at $D=\nu=0.1$. Equal values keep the two fields on the same dissipative range, so that differences between the $\phi$ and $n$ dynamics come only from the adiabatic coupling and the gradient drive. 

The adiabaticity parameter $C$ sets the strength of the parallel electron response, which couples $\phi$ and $n$ through the term $C(\phi-n)$. It is varied over $C\in[0.03,\,5]$ with logarithmic spacing. The lower end lies on the hydrodynamic side, where density and potential are weakly coupled and the dynamics approaches two-dimensional fluid turbulence. The upper end approaches the adiabatic regime, where $n\approx\phi$ and the fluctuations behave as nearly drift waves. 

The strict limit $C\to\infty$ is not considered. In this limit $n\to\phi$, the linear growth rate vanishes as $C^{-1}$, the flux $\propto\Im\{\phi_j^*n_j\}$ vanishes, and the fixed and self-consistent gradient cases become identical. The triad then reduces to the three-wave truncation of the Hasegawa--Mima equation, whose dynamics has already been characterized. Retaining the first correction in $C^{-1}$ gives the same triad with growth and damping rates $\gamma_j$, which is the model studied by Batista et al. \cite{batistaNonlinearThreemodeInteraction2006}, where these rates were introduced phenomenologically. In the present model they follow from $C$, $D$, $\nu$ and $S$. This correspondence requires $C\gtrsim20$ for the parameters used here, beyond the scanned range.

The study compares four configurations, obtained by combining two triad geometries with two treatments of the background gradient. The first comparison, between a zonal and a non-zonal triad, isolates the role of a long-wavelength mode that cannot be driven by the gradient. The second, between a fixed and a self-consistent gradient, isolates the role of the feedback of the turbulent flux on the drive.

In the self-consistent case, $S$ evolves according to Eq.~\eqref{eq:S_t} with relaxation rate $\lambda=0.5$. In the absence of fluctuations, this equation has the fixed point $S_0=P/\lambda$, so the source $P$ sets the gradient that the triad would experience without any turbulent flux. The source is scanned over $P\in[0.25,\,3]$, which corresponds to $S_0\in[0.5,\,6]$. In the fixed case, $\dot S=0$, $P$ and $\lambda$ drop out, and $S$ is scanned directly over the same interval, $S\in[0.5,\,6]$. The two maps therefore cover the same region, with $S$ in one corresponding to $S_0$ in the other. They also share the same linear stability boundary, since the condition $P>P_c=\lambda S_\mathrm{th}$ derived previously is equivalent to $S_0>S_\mathrm{th}$. Any difference between the fixed and self-consistent maps is therefore nonlinear. With $S$ fixed, the flux carried by the unstable modes has no effect on the drive, and the amplitudes saturate only through the triad transfer to damped modes. With $S$ dynamic, the flux erodes the gradient that drives it, pulling $S$ towards marginal stability.

The three wave vectors satisfy the resonance condition $\vk_1+\vk_2+\vk_3=0$, which implies the cyclic identity $\vk_1\times\vk_2=\vk_2\times\vk_3=\vk_3\times\vk_1\equiv V$. The zonal triad is
\begin{equation}
  \vk_1=(0,\,1.125),\quad \vk_2=(-0.5,\,-1.125),\quad \vk_3=(0.5,\,0),
\end{equation}
with $|k_1|=1.125$, $|k_2|\approx1.231$, $|k_3|=0.5$ and $V=0.5625$. Modes 1 and 2 are drift waves with $k_y\rho_s\sim1$, near the scale of maximal linear growth, while mode 3, with $k_{3y}=0$ and a longer wavelength, plays the role of a zonal flow. Since the drive term $-iSk_{3y}\phi_3$ vanishes, mode 3 is linearly damped for every $S$, and since its contribution to the flux is proportional to $k_{3y}\,\Im\{\phi_3^*n_3\}$, it does not enter Eq.~\eqref{eq:S_t}. It gains energy only through the triad transfer out of modes 1 and 2. The linear threshold is set by mode 1, with $S_c\approx1.74$, $0.96$, $1.13$ and $2.22$ at $C=0.03$, $0.1$, $1$ and $5$. Following the modified Hasegawa--Wakatani model\cite{numataBifurcationElectrostaticResistive2007}, the adiabatic coupling is removed for the zonal mode, since components with $k_y=0$ are constant on a flux surface and have no parallel electron response. Mode 3 is then damped only by $\nu k_3^2$ and $Dk_3^2$, and $\phi_3$ and $n_3$ decouple at linear order. Retaining $C$ for $k_y=0$ is known to artificially suppress zonal flows. Because mode 3 does not set the threshold, the stability boundary is the same in the standard and modified models.

The non-zonal triad is
\begin{equation}
  \vk_1=(0,\,1.125),\quad \vk_2=(-0.5,\,-1.0),\quad \vk_3=(0.5,\,-0.125),
\end{equation}
with $|k_1|=1.125$, $|k_2|\approx1.118$, $|k_3|\approx0.515$ and again $V=0.5625$. It is constructed to differ from the zonal triad as little as possible, keeping $\vk_1$, the coupling coefficient, and nearly the length of $\vk_3$, which is only tilted slightly off the zonal axis. All three modes are now linearly driven and all contribute to the flux. Mode 3 is weakly driven, with $S_c(\vk_3)$ increasing from $0.93$ at $C=0.03$ to $8.08$ at $C=5$. For $C\gtrsim0.065$ mode 1 remains the first to become unstable, and the stability boundary coincides with that of the zonal triad. For $C\lesssim0.065$, mode 3 becomes the most unstable mode and the threshold falls below the zonal one, down to $S_c\approx0.93$ at $C=0.03$ against $1.74$ for the zonal triad. This is the only region in which the two geometries differ already at the linear level. Elsewhere, differences between them are due to whether the long-wavelength mode can extract energy from the gradient and carry flux ($k_{3y}\neq0$) or can only receive energy through the triad transfer ($k_{3y}=0$).

The combination of the non-zonal triad with a fixed gradient deserves a separate remark. When $S$ exceeds all three thresholds $S_c(\vk_j)$, which within the scanned range occurs for $C\lesssim2.7$, every mode of the triad is linearly unstable, and no damped mode is available to absorb the energy redistributed by the nonlinear coupling. The amplitudes can then grow without bound. This cannot occur in the zonal triad, where mode 3 is always damped, nor with a self-consistent gradient, where the flux reduces $S$.

The wave modes for $\phi$ and $n$ are complex numbers, producing then a 12-dimensional phase space, and the variation of $S$ add up to 13-dimensions. Because of the frictions present, there are no conserved quantities that can be used to reduce the number of degrees of freedom, and we expect very complex behavior to be observed in the numerical simulations. 

For all the simulations, we fix the initial conditions at $\Re{\phi_j}=\Re{n_j}=0.01$ with $\Im{\phi_j}=\Im{n_j}=0$ and, in the self-consistent case, $S(0)=1$. All the equations are integrated with an adaptive fifth-order Dormand--Prince scheme\cite{dormandFamilyEmbeddedRungeKutta1980}, with absolute and relative tolerances of $10^{-6}$.  The integration runs up to $T=2.5\times10^5$.

To explore the parameter space, we compute period maps on a $1024\times1024$ grid. For a fixed gradient the grid covers the $(S,C)$ plane and for a self-consistent gradient the $(P,C)$ plane, with $S\in[0.5,6]$ or $P\in[0.25,3]$ on a linear scale and $C\in[0.03,5]$ on a logarithmic one. All points start from the initial condition given above. The asymptotic state is classified from the local maxima of an observable in the last $10^4$ time units, using $|\phi_1|$ for a fixed gradient and $S$ for a self-consistent one. A point is period-$p$ ($p\leq10$) when both the peak heights and the intervals between peaks repeat with lag $p$ over eight consecutive cycles, within a relative tolerance of $10^{-2}$. When the observable varies by less than $10^{-4}$ in the window, the point is classified as an equilibrium. Below the linear threshold of Eq.~\eqref{eq:Sc} this is the quiescent state, while above it the mode amplitudes are constant and only their phases rotate. All other points are aperiodic. Points where any amplitude exceeds $10^6$ are marked as divergent.

\section{Constant background density gradient}
\label{sec:fixed}

For comparison with a simpler case, we first consider a constant background gradient, $\dv*{S}{t}=0$. The gradient is then an external control parameter. It sets the linear drive of each mode, but the particle flux carried by the fluctuations has no effect on it. Once a mode is linearly unstable, the only mechanism that can limit its growth is the nonlinear transfer of energy to the other modes of the triad. This case also isolates the first question of this work, namely how the dynamics changes when the long-wavelength mode of the triad is zonal ($k_{3y}=0$) or non-zonal ($k_{3y}\neq0$).

\begin{figure*}[htpb]
    \centering
    \includegraphics[width=0.9\linewidth]{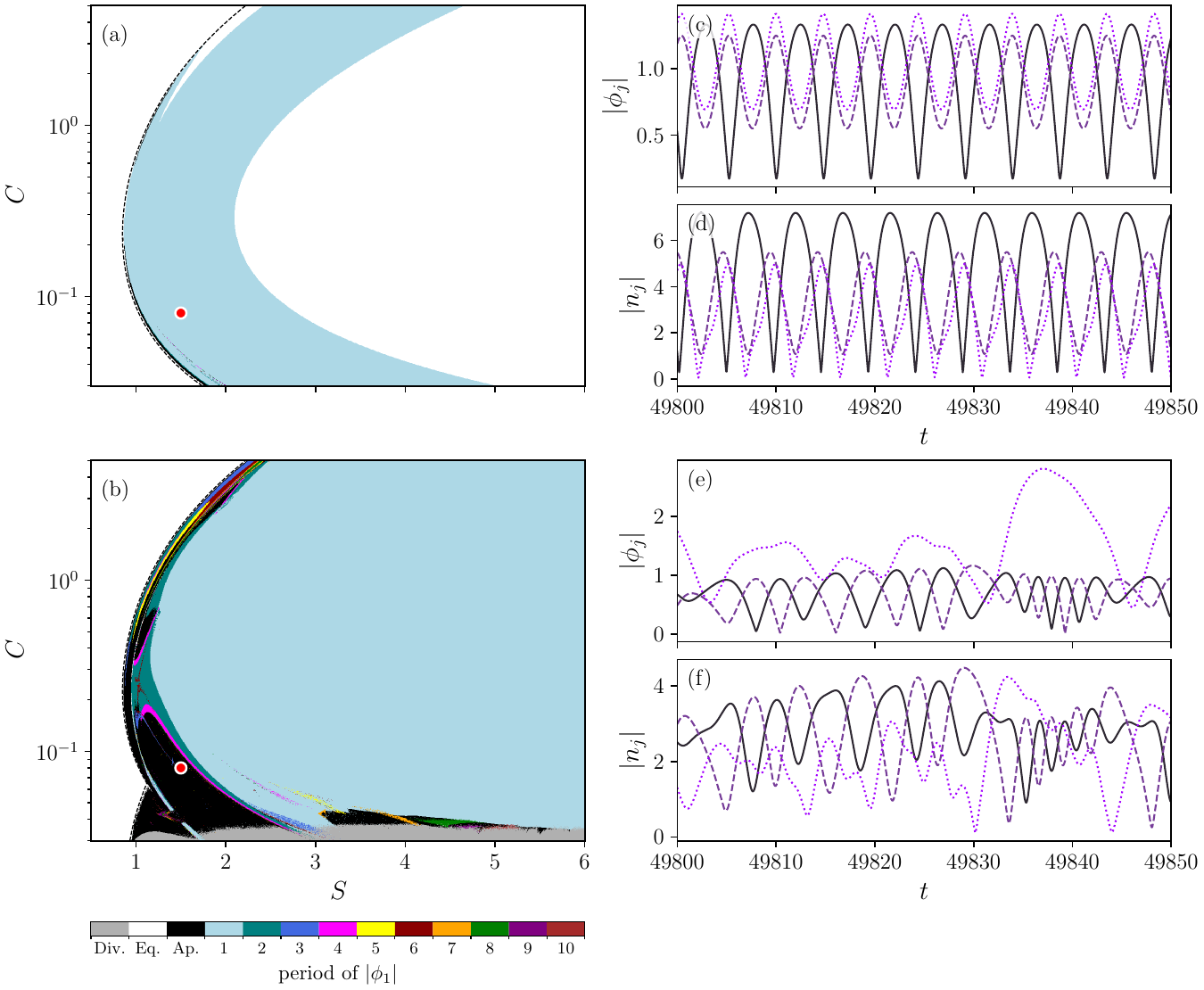}
    \caption{Dynamics with a fixed gradient $S$. (a), (b) Period of $|\phi_1|$ in the $(S,C)$ plane for the zonal and non-zonal triads, respectively. Div.: divergent; Eq.: equilibrium (constant amplitude); Ap.: aperiodic; numbers give the period. The dashed line is the linear threshold $S_\mathrm{th}$ and the red dot marks $(S,C)=(1.5,0.08)$. (c), (d) Time series of $|\phi_j|$ and $|n_j|$ for the zonal triad at the marked point. (e), (f) Same for the non-zonal triad. Solid dark line: $j=1$; purple dashed line: $j=2$; pink dotted line: $j=3$.}
    \label{fig:map_S}
\end{figure*}

Figure~\ref{fig:map_S} summarizes the results. Panels (a) and (b) are the period maps of $|\phi_1|$ in the $(S,C)$ plane for the zonal and non-zonal triads, and the dashed line is the linear threshold $S_\mathrm{th}$. We use mode 1 as the observable because it has the lowest threshold over most of the plane, so it is the first to become unstable and the main channel through which energy enters the triad. Panels (c)--(f) show the amplitudes $|\phi_j|$ and $|n_j|$ of the three modes at $C=0.08$ and $S=1.5$, the point marked in the maps.

In the zonal triad, Fig.~\ref{fig:map_S}(a), the threshold is followed by a band of period-1 motion, whose width in $S$ grows as $C$ decreases. For stronger drive the amplitudes become constant and the energy exchange between the modes stops, so these points fall in the equilibrium class, which covers most of the plane. Aperiodic motion appears only in a thin strip along the threshold, where the amplitudes are slowly modulated, and it occupies a very small part of the map. No extended aperiodic region is found for any value of $C$.

The time series at the marked point, Figs.~\ref{fig:map_S}(c) and (d), show the structure of the period-1 orbit. The amplitude of mode 1 oscillates between $0.2$ and $1.3$. Modes 2 and 3 oscillate almost in phase with each other and in antiphase with mode 1, so when $|\phi_1|$ reaches a minimum, $|\phi_2|$ and $|\phi_3|$ reach their maxima, and vice versa. This is the classical energy exchange of a resonant triad. The drift wave $\vk_1$, fed by the gradient, transfers energy to $\vk_2$ and to the zonal mode $\vk_3$, which then return it. The density follows the same pattern with larger amplitudes, $|n_j|\approx5\,|\phi_j|$, which shows that at this value of $C$ the density and the potential are far from the adiabatic response $n\approx\phi$.

This regularity is a direct consequence of the linear properties of the zonal mode. Since $k_{3y}=0$, the drive term $-iSk_{3y}\phi_3$ vanishes, and mode 3 cannot extract energy from the gradient. With the adiabatic coupling removed for $k_y=0$, it is damped only by $\nu k_3^2$ and $Dk_3^2$, and it does not contribute to the flux. Mode 3 therefore acts as a passive reservoir that can only receive energy from the drift waves and return it to them. The triad has a single source of energy, mode 1, and a well-defined path for its redistribution, and the exchange locks into a fixed cycle or into constant amplitudes.

The non-zonal triad behaves very differently, Fig.~\ref{fig:map_S}(b). Above threshold no constant-amplitude states are found, and the triad oscillates with period 1 over most of the plane. Aperiodic motion occupies a strip along the threshold for all $C$, and a broad region for $C\lesssim0.2$ and $1\lesssim S\lesssim3$, interleaved with windows of period 2 to 10. Over the whole map, a small part of the points are aperiodic. For $C\lesssim0.04$ the non-zonal triad diverges. In this hydrodynamic limit mode 3 has the lowest threshold, and all three modes become linearly unstable at moderate $S$, so no damped mode is left to absorb the energy transferred by the nonlinear coupling. In the zonal triad mode 3 is always damped, and no divergence occurs.

At the marked point, Figs.~\ref{fig:map_S}(e) and (f), the regular antiphase exchange of the zonal case is lost. Modes 1 and 2 still oscillate on the fast time scale, but with irregular amplitudes and without a fixed phase relation. Mode 3 now evolves on a slower time scale and shows large, intermittent excursions, reaching $|\phi_3|\approx2.8$, about twice the amplitude of the drift waves, before relaxing again. 

The non-zonal triad was built to differ as little as possible from the zonal one. The vector $\vk_1$ and the coupling coefficient $V$ are unchanged, the length of $\vk_3$ changes by a small factor. The only qualitative difference is that $\vk_3$ is tilted slightly off the zonal axis, which gives it a finite $k_{3y}$. This small tilt changes the role of mode 3 in two ways. It now couples linearly to the gradient, so it has its own frequency and growth or damping rate set by $S$, and it now carries part of the flux. Instead of a passive reservoir, the triad has a third active mode that competes with the drift waves for the energy of the gradient. The energy can then enter the triad through more than one mode, and the phase relation that organized the exchange in the zonal case is lost.

The onset of chaos is related to the linear thresholds of the individual modes. At $C=0.08$, mode 1 becomes unstable at $S=1.03$ in both triads. In the non-zonal triad, modes 2 and 3 follow at $S=1.13$ and $S=1.20$, and the chaotic region, $1.04\lesssim S\lesssim1.63$, starts as soon as a second mode approaches its own threshold. In the zonal triad mode 2 becomes unstable only at $S=1.52$ and mode 3 is never driven, and even with two unstable modes the motion remains regular. 

\section{Self-consistent density gradient}
\label{sec:dynamic}

\begin{figure*}[htpb]
    \centering
    \includegraphics[width=0.9\linewidth]{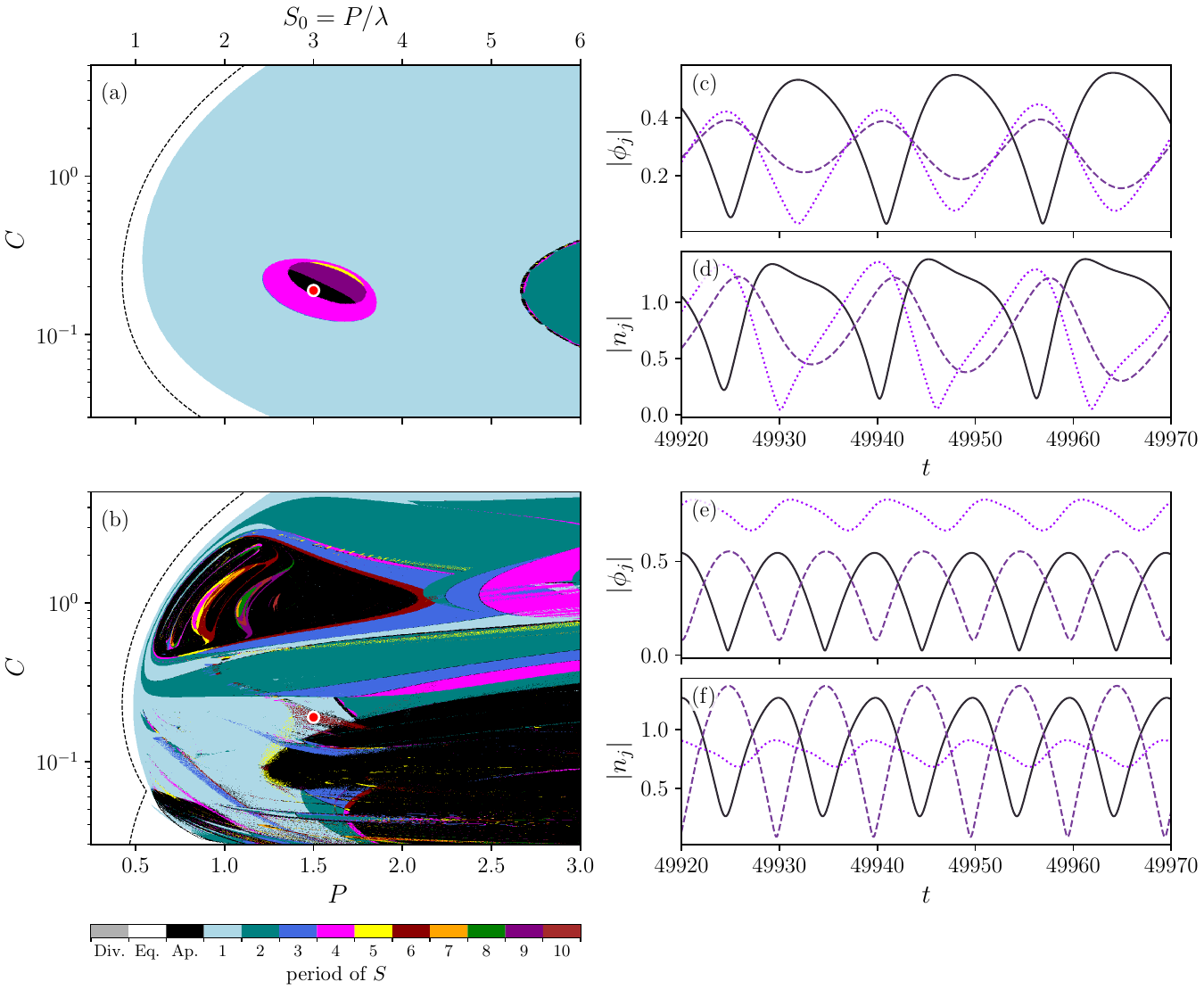}
    \caption{Dynamics with a self-consistent gradient, $\lambda=0.5$. (a), (b) Period of $S$ in the $(P,C)$ plane for the zonal and non-zonal triads, respectively; the top axis gives $S_0=P/\lambda$. Colors and dashed line as in Fig.~\ref{fig:map_S}. The red dot marks $(P,C)=(1.5,0.19)$. (c), (d) Time series of $|\phi_j|$ and $|n_j|$ for the zonal triad at the marked point. (e), (f) Same for the non-zonal triad.}
    \label{fig:map_dS}
\end{figure*}

We now let the gradient evolve according to Eq.~\eqref{eq:S_t}. $S$ is no longer a control parameter. It is fed by the source $P$, relaxes at rate $\lambda$, and is eroded by the particle flux carried by the unstable modes. This closes the loop between the drive and the fluctuations and gives the model a predator--prey structure, in which the gradient is the prey and the drift waves are the predators. Reduced models of this kind are usually written with phenomenological equations for the turbulence intensity, the zonal-flow energy and the gradient\cite{diamondSelfRegulatingShearFlow1994,kimZonalFlowsTransient2003,malkovWeakHysteresisSimplified2009}. 
In the present model the fluctuations obey the truncated Hasegawa--Wakatani equations, and the only added ingredient is the balance of Eq.~\eqref{eq:S_t}. The control parameter is now the source $P$, which sets the gradient $S_0=P/\lambda$ the plasma would have in the absence of turbulence. The top axis of Fig.~\ref{fig:map_dS} shows $S_0$, so the maps of Figs.~\ref{fig:map_S} and \ref{fig:map_dS} cover the same region.

Figure~\ref{fig:map_dS} shows the period maps of $S$ in the $(P,C)$ plane, and the amplitudes of the three modes at $C=0.19$ and $P=1.5$ ($S_0=3$). We use the line $C=0.19$ to describe the transitions in detail.

In both triads the fixed point loses stability at the threshold $P_c$ through a supercritical Hopf bifurcation, since the amplitude of the single-mode state grows as $\sqrt{P-P_c}$, Eq.~\eqref{eq:single_mode}. The threshold is followed by a thin band classified as equilibrium, in which the amplitudes are constant. It has no counterpart in the non-zonal triad with a fixed gradient. Along $C=0.19$, for $P$ just above $P_c=\lambda S_c(\vk_1)\approx0.427$, the gradient is clamped at $S=S_c(\vk_1)$ and mode 1 rotates with constant amplitude. This state is stable only in a narrow interval. Mode 3 is excited at $P\approx0.451$ in the zonal triad and at $P\approx0.434$ in the non-zonal one.  The energy taken from the gradient is now shared with the long-wavelength mode, and a steeper gradient is required to carry the same flux. The amplitudes start to oscillate at a secondary Hopf bifurcation, at $P\approx0.553$ for the zonal triad and $P\approx0.486$ for the non-zonal one, which marks the right edge of this band.

In the zonal triad, Fig.~\ref{fig:map_dS}(a), period-1 oscillations of $S$ fill most part of the map. The feedback leaves the triad regular. Aperiodic motion is present in a small portion of the parameter plane, confined to a small island around $C\approx0.2$ and $P\approx1.55$, surrounded by windows of period 4 and 9.
A second region, of period 2, appears for $P\gtrsim2.7$ and $0.1\lesssim C\lesssim0.4$, and no divergence is found.

Along $C=0.19$ the maxima of $S$ keep increasing with the source, from $S\approx1.06$ at $P=0.56$ to $S\approx1.75$ at $P=3$. They exceed the threshold of mode 2 for $P\gtrsim0.72$, but always remain far below $S_0$. The marked point lies inside the aperiodic island. The time series, Figs.~\ref{fig:map_dS}(c) and (d), show that the predator--prey cycle is superposed on the triad exchange. As with a fixed gradient, mode 1 oscillates in antiphase with modes 2 and 3, and it is the most energetic mode, with $\langle|\phi_1|^2\rangle=0.16$ against $0.09$ and $0.08$ for modes 2 and 3. The gradient, however, oscillates between $S\approx0.75$ and $1.51$. In each cycle it crosses the thresholds of both drift waves, so modes 1 and 2 are alternately driven and damped. The time average $\langle S\rangle=1.08$ is about a third of $S_0=3$. Averaged over the attractor, $\langle\Gamma\rangle=P-\lambda\langle S\rangle\approx0.96$, so roughly two thirds of the source is carried by the turbulent flux, entirely by the two drift waves, since the zonal mode carries none.

The non-zonal triad, Fig.~\ref{fig:map_dS}(b), is the most complex of the four configurations. Aperiodic motion covers a third of the map and periodic orbits of period 2 to 10 another third, while period-1 motion is reduced to half of this. The aperiodic points form two large regions. The first lies at moderate adiabaticity, $0.45\lesssim C\lesssim2.5$ and $0.65\lesssim P\lesssim2.1$, bounded by a thin layer of period 6 and surrounded by broad regions of period 2 and 3. The second lies on the hydrodynamic side, $C\lesssim0.3$, and extends to the upper edge of the drive. In contrast to the fixed gradient, no divergence is found anywhere in the map.

The finely interleaved colors in parts of the map, for instance around the marked point, do not come from the classifier. They reflect coexisting attractors, reached from the same initial condition at neighboring parameters. At $C=0.19$ and $P=1.5$, a period-1 orbit of period $T\approx9.91$ coexists with a second periodic orbit of period $T\approx47.6$, which has six maxima of $S$ and five of $|\phi_1|$ per cycle and is not a period-multiple of the first.  Their basins are intermingled and have comparable size, and which orbit is reached from a given initial condition is sensitive to perturbations of the order of the integration error.

Along $C=0.19$, after the primary instability, the gradient rises to the threshold of mode 2, which is reached at $P\approx0.48$, and then remains nearly pinned there. The profile is therefore much stiffer than in the zonal triad. The time series at the marked point, Figs.~\ref{fig:map_dS}(e) and (f), are strictly periodic. In this orbit $S$ stays below $S_c(\vk_3)$, so the long-wavelength mode is linearly damped at all times. It carries less than $2\%$ of the flux, but it holds most of the energy, $\langle|\phi_3|^2\rangle=0.57$, about four times that of each drift wave. As long as the gradient stays below $S_c(\vk_3)$, the feedback makes mode 3 behave as the zonal mode does, as a damped reservoir fed only by the triad transfer.

This changes at $P\approx1.8$. The maxima of $S$ first exceed $S_c(\vk_3)$ at $P\approx1.82$, and at almost the same point the motion becomes aperiodic. The maxima of $S$ then spread over a wide band, from $S\approx0.35$ to $2.7$ at $P=3$, and the amplitude of mode 3 reaches $|\phi_3|\approx4.5$, four times that of the drift waves. 
The mechanism is the one identified for the fixed gradient, since the triad becomes irregular when the long-wavelength mode is linearly driven. The difference is that this now happens intermittently. During each excursion of the gradient above $S_c(\vk_3)$, mode 3 is driven, grows, and releases a burst of flux that pulls $S$ back below its threshold. Because the flux always acts to reduce $S$, the amplitudes stay bounded.

The feedback does not simply suppress or enhance chaos. It moves chaos to a different region of the parameter space. With a fixed gradient, the non-zonal triad is chaotic in a narrow region just above threshold and mostly period-1 elsewhere. With a self-consistent gradient, the region near threshold has constant amplitudes or is periodic, and chaos appears at stronger drive and over a much wider range of $C$. At $C=0.08$, for instance, the fixed-gradient triad is aperiodic for $1.04\lesssim S\lesssim1.63$, whereas the self-consistent one is aperiodic for $P\gtrsim1.22$ ($S_0\gtrsim2.4$), a region that contains a locked torus and a metastable chaotic set before the chaotic attractor proper at $P\gtrsim1.32$ (Sec.~\ref{sec:routes}). Over the whole map, the fraction of aperiodic points in the non-zonal triad grows from $6\%$ with a fixed gradient to $31\%$ with a self-consistent one. In the zonal triad it remains below $1\%$ in both cases. 
The feedback therefore amplifies the difference between the two geometries. A self-consistent gradient is strongly chaotic when the long-wavelength mode can extract energy from it, and remains essentially regular when that mode is zonal.

\section{Routes to chaos}
\label{sec:routes}

Besides the regions of the parameter plane where chaos is found, it is relevant to know how it emerges to know  whether the transition is gradual or abrupt, whether it is reversible or hysteretic, and which mechanism of the model is responsible for it. In a transport problem this matters directly, because a gradual transition gives an early warning, the dynamics becomes irregular slowly, as a parameter is varied, while an abrupt one does not. We therefore follow the transitions of the non-zonal triad for both treatments of the gradient, with the adiabaticity parameter kept at $C=0.08$. The control parameter, $S$ for the fixed gradient and $P$ for the self-consistent one, is increased in small steps, each value starting from the final state of the previous one. For the self-consistent gradient the scan is repeated with $P$ decreasing, to detect coexisting attractors.

Each step is characterized by the following quantities, which are complementary. The first is the density of the maxima of an observable, $|\phi_1|$ for the fixed gradient, $S$ for the self-consistent one, which gives a bifurcation diagram, a period-$n$ orbit has $n$ distinct maxima, while a quasi-periodic or a chaotic orbit fills an interval, so this shows the structure of the attractor. The second is the power spectral density (PSD) of the same observable, which shows how many incommensurate frequencies the motion contains, a periodic orbit appears as a fundamental line with its harmonics, a quasi-periodic orbit as additional lines at combinations of two frequencies, and chaos as a continuous background. The third is the largest Lyapunov exponent $\lambda_1$, the mean exponential rate at which nearby trajectories separate, which separates regular motion ($\lambda_1=0$) from chaotic motion ($\lambda_1>0$). The phase symmetry of Eq.~\eqref{eq:symmetry} leaves two Lyapunov exponents identically zero, since rotating the phases of a solution yields another solution, so the spectrum of the full system is that of the eleven phase-invariant variables plus two zeros. The second exponent therefore carries no information and is not shown. 

These quantities say where and how the orbit changes, but not what the dynamics looks like just after chaos sets in. For that we use the laminar time. Close to some transitions the motion alternates between long, nearly regular laminar phases, during which the trajectory stays near the regular orbit that was lost, and short chaotic bursts. This is intermittency~\cite{pomeauIntermittentTransitionTurbulence1980}. The mean duration of the laminar phases, $\langle\tau\rangle$, diverges as the control parameter (here $S$) approaches its critical value $S^*$, $\langle\tau\rangle\propto S^{-\beta}$, and the exponent $\beta$ depends on the mechanism that destroys the regular orbit. In type-I intermittency, which follows a saddle-node bifurcation, $\beta=1/2$, while for types II and III, associated with a Hopf-type instability and with a period-doubling-type instability respectively, $\beta=1$. We use this analysis for two reasons. First, $\beta$ is a measurable signature of the mechanism, which allows the transition to be classified quantitatively, instead of only by eye from the bifurcation diagram and the PSD. Second, it separates soft transitions, where $\langle\tau\rangle$ diverges and the chaotic bursts appear gradually, from hard ones, where no laminar phases exist and $\langle\tau\rangle$ shows no scaling. In the present model the duration of the laminar phases is physically relevant, since during the chaotic bursts the particle flux, and therefore the gradient, changes strongly. Where the transition is hard, the relevant quantity is instead the lifetime of the chaotic transient and the width of the hysteresis loop, which we also report.

The rest of the section is organized by gradient treatment. For each triad we first look at the bifurcation diagram, the PSD and $\lambda_1$ over the whole range, then at the enlargements around each transition, where, when the transition is soft, we also use the laminar time.

\begin{figure*}[htpb]
    \centering
    \includegraphics[width=\textwidth]{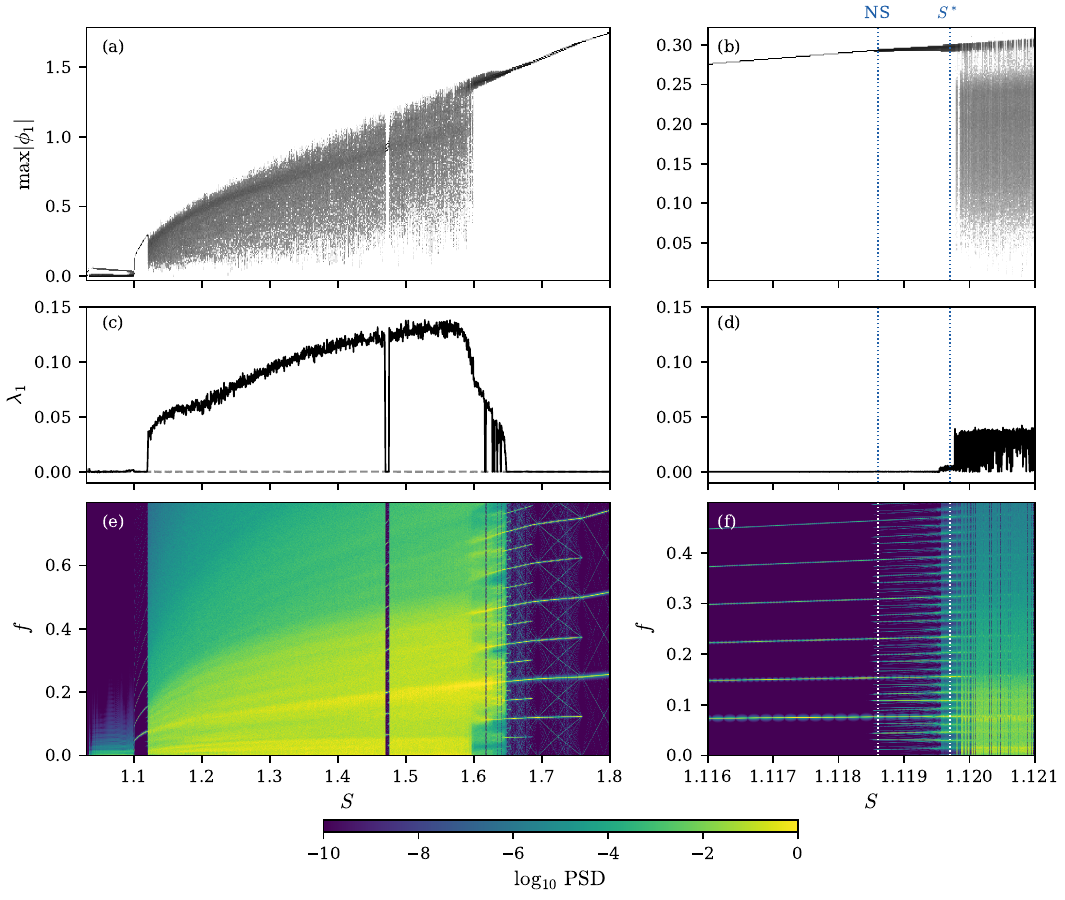}
    \caption{Route to chaos with a fixed gradient at $C=0.08$, scanned in $S$. (a), (b) Density of the maxima of $|\phi_1|$ in the non-zonal triad, with each column normalized to unit weight and shown on a logarithmic grey scale. (c), (d) Largest Lyapunov exponent $\lambda_1$ of the non-zonal (black) and zonal (dashed grey) triads. (e), (f) Power spectral density of $|\phi_1|(t)$, in $\log_{10}$ scale, as a function of frequency $f$ (Hann window, mean removed, maximum taken over frequency bins). Left: full range; right: enlargement around the quasi-periodic regime. The dotted lines in the enlargements mark the Neimark--Sacker bifurcation (NS) and the breakdown of quasi-periodicity ($S^*$).}
    \label{fig:routes_S}
\end{figure*}

\subsection{Fixed gradient}

The zonal triad has no route to chaos. Its Lyapunov exponent is zero within numerical accuracy ($\lambda_1<3\times10^{-3}$) for $1.03\leq S\leq1.8$, and after a slowly modulated state close to threshold the motion is period-1 from $S\approx1.075$.

The non-zonal triad becomes chaotic through the breakdown of quasi-periodicity, as can be seen in Fig.~\ref{fig:routes_S}. The period-1 orbit undergoes a Neimark--Sacker bifurcation at $S=1.1186$, which introduces a second, incommensurate frequency. In the period map the orbit becomes a small closed invariant curve around the fixed point. This is visible in the PSD, Fig.~\ref{fig:routes_S}(f): for $S<1.1186$ the spectrum contains only the fundamental and its harmonics, and at the bifurcation a new family of lines appears at combinations of the two frequencies, growing in amplitude with $S$. We identify the bifurcation by this appearance of a second frequency, together with the closed invariant curve in the period map, and locate it at the value of $S$ where the new lines first rise above the background. The invariant curve is destroyed at $S^*=1.1197$. At this point the lines merge into a broadband background,  $\lambda_1$ leaves zero, and the maxima fill an interval, Fig.~\ref{fig:routes_S}(b). The transition is soft. Long laminar phases, during which the trajectory stays close to the former quasi-periodic orbit, are interrupted by chaotic bursts, with a mean laminar time $\langle\tau\rangle\propto(S-S^*)^{-\beta}$ and $\beta=1.04\pm0.05$. The exponent is close to the value $1$ of type-II intermittency, which is associated with a Hopf-type instability~\cite{pomeauIntermittentTransitionTurbulence1980}.

We define the laminar time as follows. For six values of $S$ above $S^*$, logarithmically spaced in $\varepsilon=S-S^*$ from $8\times10^{-5}$ to $3\times10^{-3}$, we integrate the triad for $10^{7}$ time units after discarding a transient of $2\times10^{4}$, and record every local maximum $z_n$ of $|\phi_1|$ together with the time at which it occurs. During a laminar phase the trajectory follows the former quasi-periodic orbit, which is almost periodic, so successive maxima are nearly equal. We therefore call a maximum laminar if it differs from the next one by less than $0.02$ and lies inside the band $0.28\le z_n\le1$, which excludes the large excursions of the bursts. A laminar phase is a maximal run of consecutive laminar maxima, ended by the first maximum that fails this criterion, and only runs with at least $20$ maxima are kept. Its duration $\tau$ is the time between the first and the last maximum of the run, and phases cut by the beginning or the end of the record are discarded. The mean laminar time $\langle\tau\rangle$ at each $S$ is the average of $\tau$ over all phases found, and only values of $S$ with at least $20$ phases are used. The exponent $\beta$ is obtained from a straight-line least-squares fit of $\log\langle\tau\rangle$ against $\log\varepsilon$, with $S^*=1.11972$ kept fixed.

Chaos then persists up to $S=1.6014$, with $\lambda_1$ increasing from $0.03$ to $0.13$ and interrupted only by narrow periodic windows, which appear in the density of maxima as gaps and in the PSD as a return to sharp lines. The widest, of period 3, opens abruptly at $S=1.4690$ and coexists with the chaotic attractor at its lower edge. At $S_{\mathrm{cr}}=1.6014$ the attractor collapses in an interior crisis into a narrow chaotic band ($\lambda_1\approx0.08$). Just below $S_{\mathrm{cr}}$ the trajectory lingers in the band and escapes in bursts, as in crisis-induced intermittency~\cite{grebogiCriticalExponentsCrisisinduced1987,grebogiChaoticAttractorsCrisis1982,grebogiCrisesSuddenChanges1983}. The band is removed by a period-doubling cascade, which with decreasing $S$ reads $1\to2$ at $S=1.7587$, $2\to4$ at $1.686$, $4\to8$ at $1.6545$ and $8\to16$ at $1.6495$, and the subharmonic lines at $f_1/2^n$ are seen in the PSD. For $S>1.7587$ the motion is period-1.

\begin{figure*}[htpb]
    \centering
    \includegraphics[width=\textwidth]{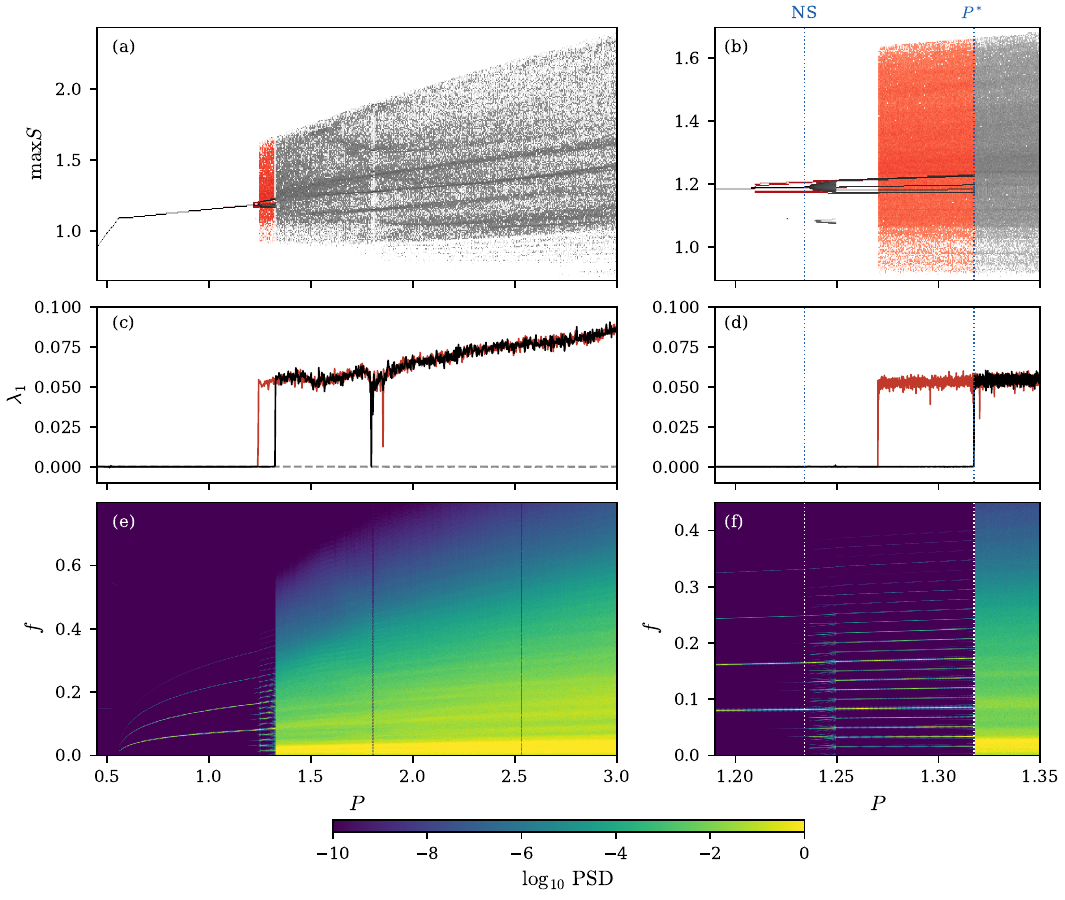}
    \caption{Route to chaos with a self-consistent gradient at $C=0.08$, scanned in $P$. (a), (b) Density of the maxima of $S$ in the non-zonal triad; the red density is the scan with $P$ decreasing and is shown only where it differs from the scan with $P$ increasing, that is, for $P<P^*$. (c), (d) Largest Lyapunov exponent $\lambda_1$, with $P$ increasing (black) and decreasing (red) for the non-zonal triad, and the zonal triad in dashed grey. (e), (f) Power spectral density of $S(t)$, as in Fig.~\ref{fig:routes_S}. Left: full range; right: enlargement around the transition, with a step of $5\times10^{-5}$ in $P$. The dotted lines mark the Neimark--Sacker bifurcation (NS) and the breakdown of the locked quasi-periodic regime ($P^*$).}
    \label{fig:routes_dS}
\end{figure*}

\subsection{Self-consistent gradient}

At the same $C$ the zonal triad is again regular. Over $0.45\leq P\leq3$ we find $\lambda_1<5\times10^{-4}$ and a periodic oscillation of $S$, with no bifurcation to chaos.

In the non-zonal triad the route has four stages, Fig.~\ref{fig:routes_dS}. Apart from the first bifurcation to a two-frequency state, they do not resemble the fixed-gradient route.

The primary Hopf bifurcation occurs at $P_c=0.515$, and the amplitudes start to oscillate at $P\approx0.557$. From there up to $P\approx1.2$ the motion is period-1 and $S$ stays between about $1.07$ and $1.19$. The maxima of $S$ form a single thin branch, and the PSD contains only the fundamental and its harmonics. The long-wavelength mode is damped at all times and the triad behaves as the zonal one. The period-1 orbit loses stability at $P_{\mathrm{NS}}\approx1.234$, where the period-1 branch ends for increasing $P$. As for the fixed gradient, we identify this as a Neimark--Sacker bifurcation from the appearance of a second frequency in the PSD and of a closed invariant curve in the period map, and we locate it at the end of the period-1 branch in the scan with step $10^{-4}$ in $P$. The rotation number lies close to $2/5$, and the motion locks into a period-5 orbit that exists for $1.249<P<1.312$, interrupted by narrow quasi-periodic intervals. In the PSD, Fig.~\ref{fig:routes_dS}(f), this stage is a comb of sharp lines at multiples of $f_1/5$, with no continuous background. The period-5 orbit persists with decreasing $P$ down to $P=1.21$.

At $P^*=1.3175$ the locked regime is replaced by a large chaotic attractor. Within a small increase in $P$, $\lambda_1$ jumps from below $10^{-3}$ to $0.05$ and the maxima of $S$ spread in the interval. The PSD changes from a line spectrum to a broadband one at the same value of $P$, with no intermediate stage in which the lines broaden or the background grows. Unlike the fixed gradient, the transition is hard. We measured the laminar time above $P^*$ as for the fixed gradient, since the regular orbit is now locked in period $5$, a maximum is laminar if it differs from the one five maxima later by less than $0.02$ and lies in the band $1.05\le z_n\le1.26$, which contains the maxima of $S$ on the locked orbit, and only runs of at least $20$ maxima are kept. Laminar phases are rare, about ten per record, and none lasts more than about $276$ time units. Their mean duration, $\langle\tau\rangle\simeq2.2\times10^{2}$, does not depend on $P-P^*$ ($\beta=0.00\pm0.01$), whereas for the fixed gradient $\langle\tau\rangle$ grows by a factor of about $50$ over a comparable range of $\varepsilon$. The laminar time therefore does not diverge at $P^*$. The transition is also hysteretic. When $P$ is decreased  the chaotic state persists down to $P=1.270$, as shown by the red branch in Fig.~\ref{fig:routes_dS}(b) and (d), where it coexists with the locked period-5 orbit reached with increasing $P$, and from the map initial condition it is still found at $P=1.2$, where it is a long chaotic transient. The lower edge of the hysteresis is therefore not sharp, and depends on the step, because the chaotic set is metastable. The chaotic set becomes permanent only when the regular orbits disappear at $P^*$. This accounts for the aperiodic region that the period map of Fig.~\ref{fig:map_dS} shows from $P\approx1.22$ at this $C$, which therefore contains a quasi-periodic orbit and a metastable chaotic set before the chaotic attractor proper.  For $1.3175<P\leq3$ the motion remains chaotic, with $\lambda_1$ increasing slowly from $0.05$ to $0.085$ and only narrow periodic windows, for instance of period 6 near $P=1.795$.

In both models the non-zonal triad reaches chaos through a Neimark--Sacker bifurcation of the period-1 orbit and the zonal triad does not. The three diagnostics agree on where each stage occurs, but they differ in what they reveal. With a fixed gradient the two-frequency regime is a visible window, $1.1186<S<1.1197$, in which new spectral lines grow before the lines merge into a continuum at $S^*$, and the breakdown is soft, through intermittency. Chaos is then a range of $S$ that closes through a crisis and period doubling. With a self-consistent gradient the two-frequency regime is largely locked in a period-5 orbit, and its breakdown at $P^*$ is abrupt and hysteretic, with no spectral precursor and no divergence of the laminar time. Chaos then has no exit within the range of drive considered.


\section{Conclusions}
\label{sec:conclusions}

We have studied a resonant triad of Hasegawa--Wakatani modes in which the background density gradient, usually imposed from outside, evolves self-consistently, fed by a particle source and eroded by the turbulent flux, so that the predator--prey loop between drive and fluctuations is closed inside the model itself. To isolate the role of the zonal flow, we compared a triad whose long-wavelength mode is zonal with one that differs only by a slight tilt of that mode, which turns it into a weakly driven drift wave, each with a fixed and with a self-consistent gradient. The linear analysis yields a closed expression for the threshold of each mode and shows that the flux of an unstable wave always lowers the gradient that feeds it. The primary instability is therefore a supercritical Hopf bifurcation, in which the gradient is clamped at the marginal stability of the first unstable mode, and the strong-adiabaticity limit recovers the triad with growth and damping rates previously introduced by hand.

The central result is the contrast between the two geometries. A zonal mode cannot extract energy from the gradient and carries no flux, so it acts as a passive reservoir, and the triad remains essentially regular for both treatments of the gradient. When the same mode is tilted off the zonal axis it competes with the drift waves for the free energy of the gradient, the orderly exchange is lost, and chaos fills a sizeable part of the parameter space. The self-consistent gradient does not simply enhance or suppress chaos. It bounds the amplitudes that diverge with a fixed gradient, moves chaos from a narrow strip near threshold to stronger drive and a wider range of adiabaticity, and greatly enlarges the fraction of aperiodic states in the non-zonal triad, so the feedback amplifies the difference between the geometries. In both cases the trigger is the same, namely the long-wavelength mode being linearly driven, which with a self-consistent gradient happens intermittently and produces bursts of flux that return the gradient to a stiff profile.

The routes to chaos of the non-zonal triad also depend on the treatment of the gradient, and none of them exists when the long-wavelength mode is zonal. With a fixed gradient, a Neimark--Sacker bifurcation is followed by a soft breakdown of the torus through intermittency, and chaos is left through an interior crisis and a period-doubling cascade. With a self-consistent gradient the torus becomes phase-locked and breaks down abruptly, with hysteresis, long metastable transients and no exit from chaos within the range of drive explored. This matters for transport, since a soft transition gives an early warning whereas an abrupt and hysteretic one does not, and it is the feedback of the flux on the gradient that makes the transition abrupt.

Taken together, these results show that a zonal mode can reduce chaos structurally, not only by shearing eddies, because it removes the competition between active modes for the free energy of the gradient, and that this protection survives when the gradient responds to the turbulence. Since the analysis is restricted to a single triad and to specific geometries, the conclusions identify a mechanism rather than a quantitative prediction for a turbulent plasma. The natural continuation is a network of coupled triads, to test whether the reduction of chaos by zonal flows and its amplification by the self-consistent gradient persist when a broad spectrum of modes interacts.

\section*{References}

\bibliography{Refs}

\end{document}